\documentstyle[aps]{revtex}

\begin{document}
\title{Nonlocality as the dynamical origin of the non-markovian process in quantum
isolated system}
\author{Hai-Jun Wang$^{*}$}
\address{Center for Theoretical Physics and School of Physics, Jilin\\
University,Changchun 130023, China}
\maketitle

\begin{abstract}
In the previous paper, it has been proved that elastic scattering processes
of two quantum particles are always accompanied with nonlocal processes.
Furthermore, it is found that setting an additional Hamiltonian after the
originally scattering one can help to describe the two type of processes in
a united frame. Here we discuss the contribution of this additional
Hamiltonian to irreversible process in isolated quantum systems. The use of
the Hamiltonian can induce the non-Markovian Langevin equation, showing a
complex memory effect, and thus revealing the irreversible essence of
isolated system without appealing to reservoir or approximate methods (e.g.
coarse grain) as usually done.

*E-mail address: whj@mail.jlu.edu.cn
\end{abstract}

The Second Law of Thermodynamics is one of fundamental laws of physics.
However, its origin remains puzzling [1]. On the macroscopic level the well
known equation $dS/dt\geq 0$ is always encountered, i.e., the processes of
an isolated system are always irreversible, whereas on the microscopic
level, the fundamental dynamical equations are all of notorious
reversibility. The puzzle arises from the difficulty of reconciling these
two conflicting sides. Inspired by the faith in macroscopic irreversibility,
many efforts have been made to explain the second law from more fundamental
points of view, such as considering fluctuation or chaos, assumption of
ergodicity from Gibbs, method of Coarse grain from von Neumann, proposing a
bath environment, the $H$-theorem from Boltzmann, star-Unitary
transformation and projective operator from Prigogine, etc. Any one of the
above assumptions can derive the macroscopic properties formally. However,
in contrast to the requirement of an irreversible theory derived from
underlying microscopic laws, rather they are\ alternative versions of
macroscopic phenomenon. From microscopic viewpoints, some other schemes,
such as the quantum measurement from Laudau and Lifshitz [2], decoherence
from Zurek [3,4] etc. are suggested, but these assertions do not pertain to
an isolated system.

To maximize the entropy is equal to approach the equivalence, say, time
arrow. The mention of time reminds us of the theory of relativity, in which
only the time interval is emphasized, but the time arrow has not been
considered. Nevertheless, by a deliberate examination one can find that in
fact, in general relativity, the dynamical equations are not reversible with
respect to time [5]. This feature brings us the hope that it is possible to
find the irreversible aspects of dynamical equations at a microscopic level.
If so, the irreversibility will automatically appear in stochastic
differential equation (e.g. Fokker-Plank equation or Langevin equation) or
evolution/transport equation (e.g. Bolzmann equation). In this paper, we
make attempt to derive an irreversible Langevin equation from fundamental
viewpoint of quantum mechanics by applying its nonlocal effect.

The model used here is simple: N fermions in an isolated box, are supposed
to be wave packages (Matter Wave) with the same initial width. After a long
time, during which many scattering processes have happened, all the
particles are expected to own a limit width of wave package in contrast to a
pure quantum wave package (without affection of any scattering) that its
width will diffuse to infinite. Through this mechanism, the wave packages
are localized and thus considered as macroscopic particles. After all, the
scattering processes with nonlocal effect may put or destroy correlations
between partial waves of the wave packages, thus destroy the regular
evolution of a pure wave package. And probably for this same factor, the
evolution equation of resultant particles is irreversible. It is interesting
to note that the localization of quantum wave is realized through nonlocal
interaction, and at the same time, the reversible microscopic dynamics for
quantum waves generates irreversible macroscopic evolving equation for
particles.

In the previous paper [6], starting from the AB-like nonlocal effect, we
have derived an additional Hamiltonian, which was demanded to contribute
only to nonlocality (geometrical phase shift). It can be assumed that the
additional Hamiltonian is too small to induce any transitions between energy
levels, as required by Berry phase[7] in adiabatic approximation. Now we
will attempt to examine this Hamiltonian by applying it to derive the
Langevin equation in order to confirm that it can induce memory effect. In
recent years, the generalized Langevin equation(GLE), i.e. equation with
memory effect, has been widely used to study non-Markovian processes in many
types of open systems[8-11]; nevertheless few of them has concerned the
isolated system (microcanonical ensemble) for the well known difficulty of
dynamical origin.

Let the additional part of Hamiltonian be denoted by ${\cal H}_N$, and the
total is 
\begin{equation}
{\cal H}={\cal H}_0+{\cal H}_N  \eqnum{1}
\end{equation}
furthermore we know that the part ${\cal H}_N$ has the same form as ${\cal H}%
_0,$ only with some coefficients before corresponding terms. The
coefficients are decided by the nonlocal region. Since ${\cal H}_0$ only
contributes to transition, and only the contribution of ${\cal H}_N$ is
meaningful for our analysis, so we just consider the part ${\cal H}_N$.
Additionally, we omit the coefficients before all of the terms of ${\cal H}_N
$ for convenience even though they may vary from term to term, the
neglecting of them will not affect our resultant analysis. In these
respects, the ${\cal H}_N$ may thoroughly possess the same form as that of $%
{\cal H}_0$, 
\begin{eqnarray}
{\cal H}_N &=&\frac 1{2m}(\vec P-e\vec A)^2+e\phi   \nonumber \\
\  &=&\frac{\vec P^2}{2m}-\frac e{2m}(\vec P\cdot \vec A+\vec A\cdot \vec P)+%
\vec A\cdot \vec A+e\phi   \eqnum{2}
\end{eqnarray}
The Hamiltonian pertains to the system that includes many particles with
both spin and charges, as mentioned in Ref.[12]. In such system the AB like
nonlocal effect always happens, in the plane where corresponding scattering
process happens, but planes are different between scattering processes, not
as required by two dimension systems. Considering the suggestion of Ref.
[12], the nonlocal effect will happen between one particle moment and
another particle's charge. However, in the general experiments on testing
the AB effect or theoretical works proposed by authors, the magnetic flux
was always assumed not to be leaking out of a certain cylinder-like region,
while in a realistic model that the flux is generated by a particle's spin $%
\vec \mu =\frac em\vec S$, the required condition is by no means preserved,
the surrounded flux is not determinable. Phenomenally, this stochastic
aspect and the scalar potential are included in the stochastic force $F(t,r)$%
, the properties of which are given only in average. 
\begin{equation}
{\cal H}_N=\frac{\vec P^2}{2m}-\frac e{2m}(\vec P\cdot \vec A+\vec A\cdot 
\vec P)+F(t,r)  \eqnum{3}
\end{equation}
where $\langle F(t,r)\rangle =0$ and $\langle F(t_1,r)F(t_2,r)\rangle $ is
known. Applying commutation relations and the Coulomb gauge $\vec \nabla
\cdot \vec A=0$, the second term of the equation can be changed to 
\begin{equation}
-\frac e{2m}(\vec P\cdot \vec A+\vec A\cdot \vec P)=-\frac em\vec A\cdot 
\vec P  \eqnum{4}
\end{equation}
To express more clearly, hereafter we use bold letter to express the vectors
to be as arguments, e.g.$\vec k\rightarrow {\bf k}$ etc. Now let us expand
the vector potential as done in quantization process [13], 
\begin{equation}
\vec A({\bf x},t)=\sum_{{\bf k}}\sum_r(\frac{\hbar c^2}{2V\omega _{{\bf k}}}%
)^{\frac 12}\vec \varepsilon _r({\bf k})[a_r({\bf k})e^{i({\bf k}\cdot {\bf x%
}-\omega t)}+a_r^{\dagger }({\bf k})e^{-i({\bf k}\cdot {\bf x}-\omega t)}] 
\eqnum{5}
\end{equation}
and when $t=0$, the corresponding vector potential is 
\begin{equation}
\vec A({\bf x})=\sum_{{\bf k}}\sum_r(\frac{\hbar c^2}{2V\omega _{{\bf k}}})^{%
\frac 12}\vec \varepsilon _r({\bf k})[a_r({\bf k})e^{i{\bf k}\cdot {\bf x}%
}+a_r^{\dagger }({\bf k})e^{-i{\bf k}\cdot {\bf x}}]\;\text{.}  \eqnum{6}
\end{equation}
Substitute the above expansion into eq. (5), 
\begin{equation}
-\frac em\vec A\cdot {\bf p}=-\frac em\sum_{{\bf k}}\sum_r(\frac{\hbar c^2}{%
2V\omega _{{\bf k}}})^{\frac 12}\vec \varepsilon _r({\bf k})\cdot {\bf p}%
[a_r({\bf k})e^{i{\bf k}\cdot {\bf x}}+a_r^{\dagger }({\bf k})e^{-i{\bf k}%
\cdot {\bf x}}].  \eqnum{7}
\end{equation}
Now starting from the above equation, let us discuss the dipole
approximation and quadrupole approximation respectively. Although the
terminology is quite similar to the calculation in the transition amplitude,
but as mentioned above, the nonlocal Hamiltonian will not induce any
transition. We only use the approximation as a tool, bearing in mind that
the Hamiltonian used here is only the part of nonlocality.

{\it (i) Dipole approximation} While $\lambda \gg \mid \vec x\mid $, ${\bf k}%
\cdot {\bf x}\ll 1$, $e^{i{\bf k}\cdot {\bf x}}\sim e^{-i{\bf k}\cdot {\bf x}%
}\sim 1$, then the equation (8) yeilds 
\begin{equation}
-\frac em\vec A\cdot {\bf p}=\sum_{{\bf k},r}V_{{\bf k},r}({\bf p})[a_r({\bf %
k})+a_r^{\dagger }({\bf k})]  \eqnum{8}
\end{equation}
where $V_{{\bf k},r}({\bf p})=-\frac em(\frac{\hbar c^2}{2V\omega _{{\bf k}}}%
)^{\frac 12}\vec \varepsilon _r({\bf k})\cdot {\bf p}$. And the Hamiltonian
in eq.(4) now reads 
\begin{equation}
{\cal H}_N=\frac{{\bf p}^2}{2m}+F(t,r)+\sum_{{\bf k}}\hbar \omega _{{\bf k}%
}a_r^{\dagger }({\bf k})a_r({\bf k})+\sum_{{\bf k},r}V_{{\bf k},r}({\bf p}%
)[a_r({\bf k})+a_r^{\dagger }({\bf k})].  \eqnum{9}
\end{equation}
the additional terms of $\hbar \omega _{{\bf k}}a_r^{\dagger }({\bf k})a_r(%
{\bf k})$ have derived from the pure energy of the electromagnitic energy $%
\vec E^2+\vec B^2$. In following let us evaluate the evolving equation for
coordinates, momentum, $a_r({\bf k})$ and $a_r^{\dagger }({\bf k})$
respectively: 
\begin{equation}
{\bf \dot x=}\frac i\hbar [{\cal H}_N,{\bf x}]=\frac{{\bf p}}m+\sum_{{\bf k}%
,r}\bigtriangledown _{{\bf p}}V_{\vec k,r}({\bf p})[a_r({\bf k}%
)+a_r^{\dagger }({\bf k})]  \eqnum{10}
\end{equation}
where we have used the commutation $[f({\bf p}),{\bf x}]=-i\hbar \vec 
\bigtriangledown _{{\bf p}}f({\bf p})$. For momentum, 
\begin{equation}
{\bf \dot p=}\frac i\hbar [{\cal H}_N,{\bf x}]=0  \eqnum{11}
\end{equation}
and 
\begin{eqnarray}
\dot a_r^{\dagger } &=&\frac i\hbar [{\cal H}_N,a_r^{\dagger }]=i\omega _{%
{\bf k}}a_r^{\dagger }+\frac i\hbar V_{{\bf k},r}({\bf p})  \eqnum{12a} \\
\dot a_r &=&\frac i\hbar [{\cal H}_N,a_r^{\dagger }]=-i\omega _{{\bf k}}a_r-%
\frac i\hbar V_{{\bf k},r}({\bf p})  \eqnum{12b}
\end{eqnarray}
The equation (13a) and (13b) can be directly resolved by applying Laplace
transformation. Since the two equations are Hermite to each other, only one
of them is necessarily considered. The result for $a_r^{\dagger }(t)$ reads 
\begin{equation}
a_r^{\dagger }(t)=a_r^{\dagger }(0)e^{-i\omega _{{\bf k}}t}+\frac i\hbar
\int_0^te^{-i\omega _{{\bf k}}\tau }V_{{\bf k},r}({\bf p}(t-\tau ))d\tau 
\eqnum{13}
\end{equation}
As $V_{{\bf k},r}({\bf p}(t-\tau ))$ is independent of time $t$, the
integration can be carried out, 
\begin{eqnarray}
a_r^{\dagger }(t) &=&a_r^{\dagger }(0)e^{-i\omega _{{\bf k}}t}+\frac i\hbar
V_{{\bf k},r}({\bf p})\int_0^te^{-i\omega _{\vec k}\tau }d\tau  \nonumber \\
\ &=&a_r^{\dagger }(0)e^{-i\omega _{{\bf k}}t}+\frac{V_{{\bf k},r}({\bf p})}{%
\hbar \omega _{{\bf k}}}[1-e^{-i\omega _{{\bf k}}t}]  \eqnum{14}
\end{eqnarray}
Applying the result to equation (11), one obtains 
\begin{eqnarray}
{\bf \dot x} &&{\bf =}\frac{{\bf p}}m+\sum_{{\bf k},r}\vec \bigtriangledown
_{{\bf p}}V_{{\bf k},r}({\bf p})[a_r({\bf k})+a_r^{\dagger }({\bf k})] 
\nonumber \\
\ &=&\frac{{\bf p}}m+\sum_{{\bf k},r}V_0({\bf p})[a_r({\bf k})+a_r^{\dagger
}({\bf k})]\vec \varepsilon _r({\bf k})\cdot (\vec \bigtriangledown _{{\bf p}%
}{\bf p})  \eqnum{15}
\end{eqnarray}
where $V_0({\bf p})=(-\frac em)(\frac{\hbar c^2}{2V\omega _{{\bf k}}})^{%
\frac 12}$. It can be noticed that in this case there appears no memory
effect, so the dipole process from nonlocality has nothing to do with the
non-Markovian process. Then let us turn to quadrupole process.

(ii) {\it Quadrupole approximation:} $e^{i\vec k\cdot \vec x}\sim 1+i{\bf k}%
\cdot {\bf x},$ $e^{-i\vec k\cdot \vec x}\sim 1-i{\bf k}\cdot \vec x$. If
omitting the interaction terms that appeared in dipole approximation, the
qualitative results will not be affected. By doing so and employing the same
procedure as the above equations, the equation (7) yields 
\begin{equation}
-\frac em\vec A\cdot {\bf p}=\sum_{{\bf k},r}V_{{\bf k}}({\bf p},{\bf x}%
)[a_r({\bf k})-a_r^{\dagger }({\bf k})]  \eqnum{16}
\end{equation}
where $V_{{\bf k}}({\bf p},\vec x)=i(-\frac em)(\frac{\hbar c^2}{2V\omega _{%
{\bf k}}})^{\frac 12}\vec \varepsilon _r({\bf k})\cdot {\bf pk}\cdot {\bf x}$%
. And the nonlocal Hamiltonian turns to 
\begin{equation}
{\cal H}_N=\frac{{\bf p}^2}{2m}+F(t,r)+\sum_{{\bf k},r}\hbar \omega _{{\bf k}%
}a_r^{\dagger }({\bf k})a_r({\bf k})+\sum_{{\bf k},r}V_{{\bf k},r}({\bf p},%
{\bf x})[a_r({\bf k})-a_r^{\dagger }({\bf k})]  \eqnum{17}
\end{equation}
Again, calculating the time derivatives of coordinate and momentum, $a_r(%
{\bf k})$ and $a_r^{\dagger }({\bf k})$: 
\begin{equation}
{\bf \dot x=}\frac i\hbar [{\cal H}_N,{\bf x}]=\frac{{\bf p}}m+\sum_{{\bf k}%
,r}\vec \bigtriangledown _{{\bf p}}V_{{\bf k},r}({\bf p,x})[a_r({\bf k}%
)-a_r^{\dagger }({\bf k})]  \eqnum{18}
\end{equation}
\begin{equation}
{\bf \dot P=}\frac i\hbar [{\cal H}_N,{\bf p}]=-\sum_{{\bf k},r}\vec 
\bigtriangledown _{{\bf x}}V_{{\bf k},r}({\bf p,x})[a_r({\bf k}%
)-a_r^{\dagger }({\bf k})]  \eqnum{19}
\end{equation}
where we have used the commutator $[{\bf p},f({\bf x)}]=-i\hbar \vec 
\bigtriangledown _{{\bf x}}f({\bf x})$, and 
\begin{eqnarray}
\dot a_r^{\dagger } &=&\frac i\hbar [{\cal H}_N,a_r^{\dagger }]=i\omega _{%
{\bf k}}a_r^{\dagger }+\frac i\hbar V_{{\bf k},r}({\bf p})  \eqnum{20a} \\
\dot a_r &=&\frac i\hbar [{\cal H}_N,a_r^{\dagger }]=-i\omega _{{\bf k}}a_r+%
\frac i\hbar V_{{\bf k},r}({\bf p})  \eqnum{20b}
\end{eqnarray}
the eqs.(20) can be resolved by using the Laplace transformation like eqs.
(13). Since the above two equations are not Hermite to each other, the
solutions should be written out separately 
\begin{equation}
a_r^{\dagger }(t)=a_r^{\dagger }(0)e^{-i\omega _{{\bf k}}t}+\frac i\hbar
\int_0^te^{-i\omega _{{\bf k}}\tau }V_{{\bf k},r}({\bf p}(t-\tau ),{\bf x}%
(t-\tau ))d\tau  \eqnum{21}
\end{equation}
\begin{equation}
a_r(t)=a_r(0)e^{i\omega _{{\bf k}}t}+\frac i\hbar \int_0^te^{i\omega _{{\bf k%
}}\tau }V_{{\bf k},r}({\bf p}(t-\tau ),{\bf x}(t-\tau ))d\tau  \eqnum{22}
\end{equation}
However, the integrals can't be carried out, because $V_{{\bf k},r}$ is of
intricate time-dependence due to that the momentum in it is determined by
eq.(19), which is not trivial now. Substitute the above solution to eq.(18)
and eq.(19), one obtains 
\begin{equation}
{\bf \dot x=}\frac{{\bf p}}m-\frac{2i}\hbar \sum_{{\bf k},r}V_{{\bf 0}}({\bf %
k})\vec \varepsilon _r({\bf k}){\bf k}\cdot {\bf x}\int_0^t\sin \omega \tau
V_{{\bf k},r}(t-\tau )d\tau +\tilde F_q(t)  \eqnum{23}
\end{equation}
\begin{equation}
{\bf \dot p=}\frac{2i}\hbar \sum_{{\bf k},r}V_0({\bf k})\vec \varepsilon _r(%
{\bf k})\cdot {\bf p\;k}\int_0^t\sin \omega \tau V_{{\bf k},r}(t-\tau )d\tau
+\tilde F_p(t)  \eqnum{24}
\end{equation}
where the random forces are 
\begin{equation}
\tilde F_q(t)=\sum_{{\bf k},r}iV_{{\bf 0}}({\bf k})\vec \varepsilon _r({\bf k%
}){\bf k}\cdot {\bf x}[a_r(0)e^{i\omega _{{\bf k}}t}-a_r^{\dagger
}(0)e^{-i\omega _{{\bf k}}t}]  \eqnum{25}
\end{equation}
\begin{equation}
\tilde F_p(t)=-\sum_{{\bf k},r}iV_{{\bf 0}}({\bf k})\vec \varepsilon _r({\bf %
k})\cdot {\bf p}\;{\bf k}[a_r(0)e^{i\omega _{{\bf k}}t}-a_r^{\dagger
}(0)e^{-i\omega _{{\bf k}}t}]  \eqnum{26}
\end{equation}
This type of random forces is determined by the initial condition. The
original force seems vanished but in fact it has affection on the $a_r(0)$
and $a_r^{\dagger }(0)$ as an initial stochastic input. The present
stochastic force decided directly by initial condition and thus are relevant
to the original stochastic force $F(r,t)$. Anyway, the time-dependence of
eq.(23) and eq.(24) is obvious and unavoidable.

It is difficult to resolve the equations (23) and (24) as they are of
intricate manner of self-dependence and nonlinearity. As for the eq.(24), it
is still not easy to treat the integral even with the approximation of
making the outside part of the integral sign as a constant vector, because
the Laplace transformation can't be performed on the integrand $V_{{\bf k}%
,r}(t-\tau )$ that includes all the components of ${\bf p}(t-\tau )$ as $%
\vec \varepsilon _r({\bf k})\cdot {\bf p}(t-\tau )\,{\bf k}\cdot {\bf x}%
(t-\tau )$. So it is impossible to treat the equation as general Langevin
equation: $m\dot \upsilon (t)=-U^{\prime }(x)-\int_0^t\beta (t-t^{\prime
})\upsilon (t^{\prime })dt^{\prime }+\varepsilon (t)$ [8], to which the
method of Laplace transformation can be applied. So the solution of these
equations deserve further research.

{\it Conclusion and Discussion }Conventionally we hold that if the number of
photons is stable: $\langle \sum_{i=1}\hbar \omega _{\vec k}a^{\dagger
}a\rangle =constant$, then the systems possess the equilibrium properties.
But essential aspect of recognizing the stochastic in a quantum system
should not be confined to the knowledge of energy levels. It is emphasized
here that the random phase of the wave function should also be considered
properly. And being included in the dynamical equation, the changes of phase
may be helpful in treating the reversible puzzle of fundamental dynamical
equations. We notice from the eq.(23) and eq.(24) that random phases are not
of a purely stochastic character, with its variation constrained by the
dynamical equation as follows. If assuming that the particle starts at time $%
t=0$ and at $x=x_0$ with the velocity $\upsilon =\upsilon _0$, the mean
square value of its displacement at time $t$ is given by [14], 
\begin{equation}
\langle (x(t)-x_0)^2\rangle =\int_0^t\int_0^t\langle \upsilon (t_1)\upsilon
(t_2)\rangle dt_1dt_2  \eqnum{27}
\end{equation}
where $\upsilon (t)$ is ${\bf \dot x}$ in eq.(23) and the ${\bf p}$ in it is
resolved from eq.(24). The equation (27) adds a strong constraint to any
wave package's evolution, for example, to Gauss wave package. The equation
will modify the evolution of Gauss wave package out of regular way and thus
damage the diffusion mechanism and coherence state of constituent plane
waves[3, 4, 15]. In this sense, the memory effect just exhibits its effects
by damaging the regular evolution of quantum wave, which is originally a
reversible evolution.

The reversible evolution without nonlocality can actually be seen from the
Feynman path integral method for a quantum wave: the form $\psi ({\bf x}%
_2,t_2)=\int K({\bf x}_2,t_2;{\bf x}_1,t_1)\psi ({\bf x}_1,t_1)d^3{\bf x}_1$
is obviously Markovian, because all information of any moment can induce the
wave of any other moment. The Schr\"odinger equation possesses the same
essence, as we know, it is of the same virtue as the Feynman method in
describing quantum mechanics. We conclude here that these reversible form
can be broken if the nonlocal effect is introduced.

The motivation of this paper arose from the relationship of two unknown: the
gauge field born from unknown, and where the unknown exists is just the
place we use statistics and stochastic.

In conclusion, this paper only resolves the problem how the quantum process
is irreversible due to the quantum nonlocal effect. However, whether the
system will evolve into equilibrium is not known for the absence of the
exact solutions to equations (23) and (24). Therefore to understand how the
system approaches the equilibrium, the study on the solution of the
equations is of importance. It may evolve into equilibrium under some
situation, and may not for other cases. Peculiarly, some initial conditions
may also be critical for the solution, e.g. the stochastic force origin, the
nonequilibrium distribution of particles, etc.

\end{document}